\documentclass{article}
\usepackage{amsmath}
\usepackage{latexsym}
\usepackage{makeidx}
\usepackage{natbib}
\usepackage{verbatim}
\usepackage{graphicx}
\usepackage{doi}
\usepackage{pdfpages}
\usepackage{float}
\begin{document}

  \title{\hspace{1em} A note on goodness of fit testing \newline for the Poisson distribution}
  \author{Erekle  Khurodze\footnote{V Chavchanidze Institute of Cybernetics, Georgian Technical University.\newline \mbox{\hspace{1.5em}} Email: e.khurodze2@gmail.com.}, \hspace{.3em}Leigh A Roberts\footnote{Email: leighroberts745@gmail.com (corresponding author).}}
\maketitle

\noindent\small {\bf Abstract}

Since its introduction in 1950, Fisher's dispersion test has become a standard means of deciding whether or not count data follow the Poisson distribution.
The test is based on a characteristic property of the Poisson distribution, and discriminates well between the Poisson and the natural alternative hypotheses of binomial and negative binomial distributions.

While the test is commonly used to test for general deviations from Poissonity, its performance against more general alternatives has not been widely investigated.
This paper presents realistic alternative hypotheses for which general goodness of fit tests perform much better than the Fisher dispersion test.

\bigskip

\noindent\\
{\it Key words.} Poisson, binomial, negative binomial distributions; test of Poissonity; Fisher index of dispersion; goodness of fit statistic; power of test; normal, gamma, Weibull distributions.\\
{\it Acknowledgement.} Thanks to Estate Khmaladze for
helpful comments.

\bigskip\normalsize

\section*{Introduction}

\hspace{1.5em}Given the tractability and ease of interpretation of the Poisson distribution,
a natural first step in the analysis of count data is to decide whether or not the data could reasonably arise from a Poisson generating process.

Should this test fail, a logical second step is to consider whether the data could arise from binomial or negative binomial distributions.
These distributions are related to the Poisson distribution, and are distinguished from it by under- and over-dispersion: the binomial has variance less than the mean, while the variance of the negative binomial exceeds the mean; and for the Poisson they are equal.  

The aim of this paper is to illustrate that a particular test statistic, designed in the light of its relationships within the Poisson/binomial/negative binomial family of distributions, may be powerful in selecting binomial and negative binomial alternatives when the Poisson null is incorrect; but that same statistic may be less effective against more general alternative hypotheses.

The test statistic in question is essentially the Fisher index, which seems to be regarded as an omnibus test for Poissonity.  Alternative hypotheses considered include mixtures of the binomial distribution on the one hand, and discrete data arising as truncated continuous random variables on the other.  The continuous variables follow the normal, gamma and Weibull distributions, and all the alternative hypotheses could plausibly arise in practice, and are of interest to practitioners.

A defining characteristic of the Poisson distribution is equality of the mean and variance, and alternative hypotheses investigated here are chosen to have equal mean and variance.

The structure of the paper is as follows.
In \S\S 1 and 2 we discuss respectively the $\hat c$ test for the Poisson distribution, and the goodness of fit (gof) tests that we intend to apply.  After clarifying the nature of the numerical simulations in \S3, we consider experiments within the `family' of the Poisson, binomial and negative binomial distributions.  The following section concerns testing when alternative hypotheses are drawn from outside this family; and we summarise in a short conclusion.

\section{The $\hat c$ test\label{s1}}

\hspace{1.5em}In an encyclopedic survey of tests for whether data could be generated by a Poisson mechanism,
\cite{mijburgh-visagie2020a}
give an overview of the literature, and provide estimates of power for a variety of test statistics against a raft of competing hypotheses using bootstrap methods.

In contrast, the present paper
considers only three test statistics and a limited number of alternative hypotheses.  
We compare empirical distribution functions (EDFs)
of these statistics
under contrasting pairs of generating mechanisms, and evaluate their relative power.

Perhaps the most basic test for Poissonity is Fisher's variance test, or index of dispersion, or simply Fisher's index, equal to the sample variance divided by the mean \citep{fisher1950a}.  Denoting this ratio by $FI$, when $FI-1$ is not sufficiently close to zero to allow one to claim that the data follows a Poisson distribution, there is an implied preference for the binomial or negative binomial.

In testing null Poisson hypotheses we work with a rescaled version of $FI-1$.  Our starting point is \cite{kyriakoussis-li-papadopoulos1998a}, in which the relationships between the Poisson, binomial and negative binomial distributions is explored.
The authors consider a distribution on the non-negative integers with a probability mass function (pmf) of the form
\begin{equation}
p_\theta(x)=\frac{\alpha(x)\,\theta^x}{\sum_{j=0}^M\,\alpha(j)\,\theta^j}\qquad\mbox{for}\ x=0,1,\ldots,M
\label{e1}
\end{equation}
where $M$ is a positive integer or infinity;
$\theta$ is an unknown parameter assuming values in an open set $\Theta=(0,L)$ for positive $L$; and
$\alpha(x)>0$ for $x=0,1,\ldots,M$.

Considering a random variable $X$ with pmf given by \eqref{e1}, with finite second moment, they define
\[
c=\frac{E_\theta(X(X-1))}{[E_\theta X]^2}
\]
and prove that $X$ has the Poisson, binomial or negative binomial distribution if and only if, for all $\theta\in\Theta$, the value of $c$ is $1,\frac{m-1}m,\frac{m+1}m$ respectively, for a positive integer $m$.
This integer has a direct interpretation:
for the binomial $m$ denotes the number of trials, while for the negative binomial
we count the number of failures occurring until the $m$
th success turns up.  In the remainder of the paper these parameters are denoted respectively by $m_b$ and $m_{nb}$.

The sample analogue of $c$ is then defined, viz.:
\[
\hat c=\frac{\frac1n\sum_{i=1}^nX_i(X_i-1)}{(\frac1n\sum_{i=1}^nX_i)^2}
\]
whence, defining $\bar X=\frac1n\sum_{i=1}^nX_i$ and $s^2=\frac1n\sum_{i=1}^n(X_i-\bar X)^2$, we have
\[
\hat c-1=\frac{s^2-\bar X}{\bar X^2}=\frac{FI-1}{\bar X}
\]
The authors then show that, correcting for the mean and normalising, $\hat c$ has a limiting normal distribution.  In the case of the Poisson distribution, the limiting distribution assumes a simple form:
\begin{equation}
  \frac{\sqrt n(\hat c-1)}{
    \sqrt2/\bar X
  }\overset{d}\to{\cal N}(0,1)\qquad\mbox{as}\quad n\to\infty
\label{e3}
\end{equation}
\citep[Thm.\ 2]{kyriakoussis-li-papadopoulos1998a}.  Analogous limiting normal distributions for the binomial and negative binomial distributions are more complicated.

In this paper, we use the numerator of the left side of \eqref{e3}, viz.\ $\sqrt n(\hat c-1)$, to test the null hypothesis of a Poisson distribution (referred to as the `$\hat c$ test'), and compare the  power of this test with gof tests under various scenarios.
We test initially within the family of the three distributions, and then against a limited number of alternative hypotheses.

\section{Goodness of fit testing}

\hspace{1.5em}
The gof test assumes the following form. First we group
possible observations
into $K$ cells, with the delineation of cells undertaken prior to the experiment; for a sample of $\{X_i\}_{i=1}^n$ denote by $\nu_k$ the number of observations belonging to the $k$th cell, and by $p_k$ the probability of an observation falling into the $k$th cell, calculated according to the null or fitted probability distribution.

Setting
\begin{equation}
  Y_k(\theta)=\frac{\nu_k-np_k(\theta)}{\sqrt{np_k(\theta)}}\qquad\mbox{for}\quad k=1,2,\ldots,K
\label{e4}
\end{equation}
the gof test statistic is defined as
\begin{equation}
T(\theta)=\frac1{\sqrt K}\max_k\left|\sum_{k^{\,\prime}\leq k}Y_{k^{\,\prime}}(\theta)\right|
\label{e2}
\end{equation}
where $\theta$ is either a value $\theta_0$ fixed by the null hypothesis, or the maximum likelihood estimator (MLE) $\hat\theta$ of the unknown parameter of the family fixed by the null hypothesis.  Since we are only testing Poisson null hypotheses, the probabilities in \eqref{e4} are calculated from the Poisson pmf, and the MLE is the sample mean.

The only grouping of the data in the calculation of $T(\theta)$ undertaken in this paper is truncation from above or below.
With each gof test of a null, values of $k_{max}$ and $k_{min}$ are given.
Any observed value exceeding $k_{max}$ is placed in the same cell as $k_{max}$, which is the least member of that cell; and analogously when an observation falls below a lower limit of $k_{min}$.  The probabilities of an observation falling in an extreme cell are respectively $p_{max}$ and $p_{min}$.

There is no grouping of data between these two barriers, and calculation of $\hat c$ and the sample mean are unaffected by any grouping, since they use the original sample values.

Subject to a minimum expected number of observations in each cell of say 4 to 6, the limiting distribution of $T$ for large $K$ is known; but with the modest values of $K$ in this paper we work with Monte Carlo simulations rather than limiting distributions.

\section{Nature of the experiments}

\hspace{1.5em}
Each experiment has two halves, the first half of which generates a sample of $n$ Poisson random variables, with parameter $\lambda$; and the second of which generates a sample of size $n$ from an alternative distribution.

To each half of the experiment we apply three statistical tests for a Poisson null: first comes the $\hat c$ test, as described above in \S\ref{s1}, to be followed by two applications of \eqref{e2}.  The parameter $\theta$ in the statistic $T(\theta)$ is initially set at $\lambda$, following which it is given the value of the MLE.

Results of each experiment are summarised in a graph, with the captions showing the values of $k_{min},k_{max},p_{min}$ and $p_{max}$, together with parameters defining the alternative distribution being applied.
Continuous lines in the figures indicate the EDFs found when applying the tests to the first sample, when the null Poisson hypothesis is in fact true; and broken lines show the results of the same statistical tests applied to the second half of the experiment, when the data is generated according to an alternative hypothesis.

These alternative distributions are initially the binomial distribution and the negative binomial distribution, staying within the `family' of the three basic distributions for count data.  Moving away from this family, we consider in turn the beta binomial distribution; a particular mixture of a binomial and a negative binomial; and finally three continuous distributions, with disrete observations obtained from rounding a continuous random variable.
These last three are the normal, gamma and Weibull distributions.

Throughout the paper, the number of replications in each experiment is 5000; and the sample size is 100 for the first six experiments, 200 for the final three.

\begin{figure}[htbp]\centering
  \includegraphics[width=0.6\textwidth,height=0.5\textwidth]{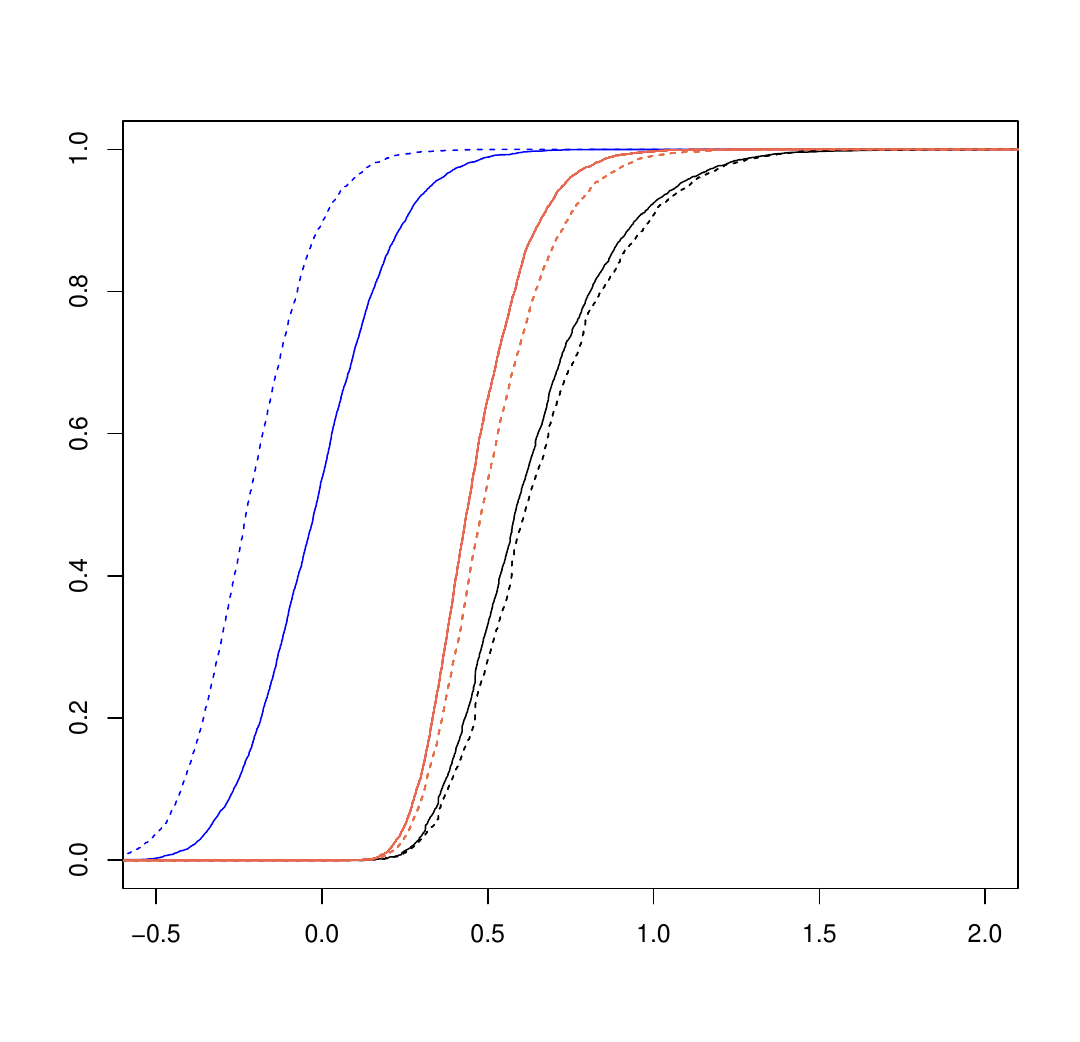}
  \caption{Binomial Alternative, with $k_{min}=3,k_{max}=12,p_{min}=0.082,p_{max}=0.053$; and $m_b=50,p_b=0.14$.}
\label{erekl1}
\end{figure}

\section{Testing within the three basic count distributions}

\hspace{1.5em}
The first alternative hypothesis is the binomial distribution.  Parameter values are $m_b=50,p_b=0.14$, so that the mean is in agreement with that of the null Poisson distribution with parameter $\lambda=7$.  The variance of this binomial distribution is 6.0.

Results from the first experiment are shown in Fig \ref{erekl1}.  It is seen that the $\hat c$ test shows significantly greater power than the gof tests based on \eqref{e2}.  The values of $p_{min}$ and $p_{max}$ are calculated from the Poisson distribution with parameter $\lambda$.  These are correct for the first gof test using \eqref{e4}, but should also be close to the actual values when the MLE $\hat\theta$ is used in the second gof test.

\begin{figure}[htbp]\centering
\includegraphics[width=0.6\textwidth,height=0.5\textwidth]{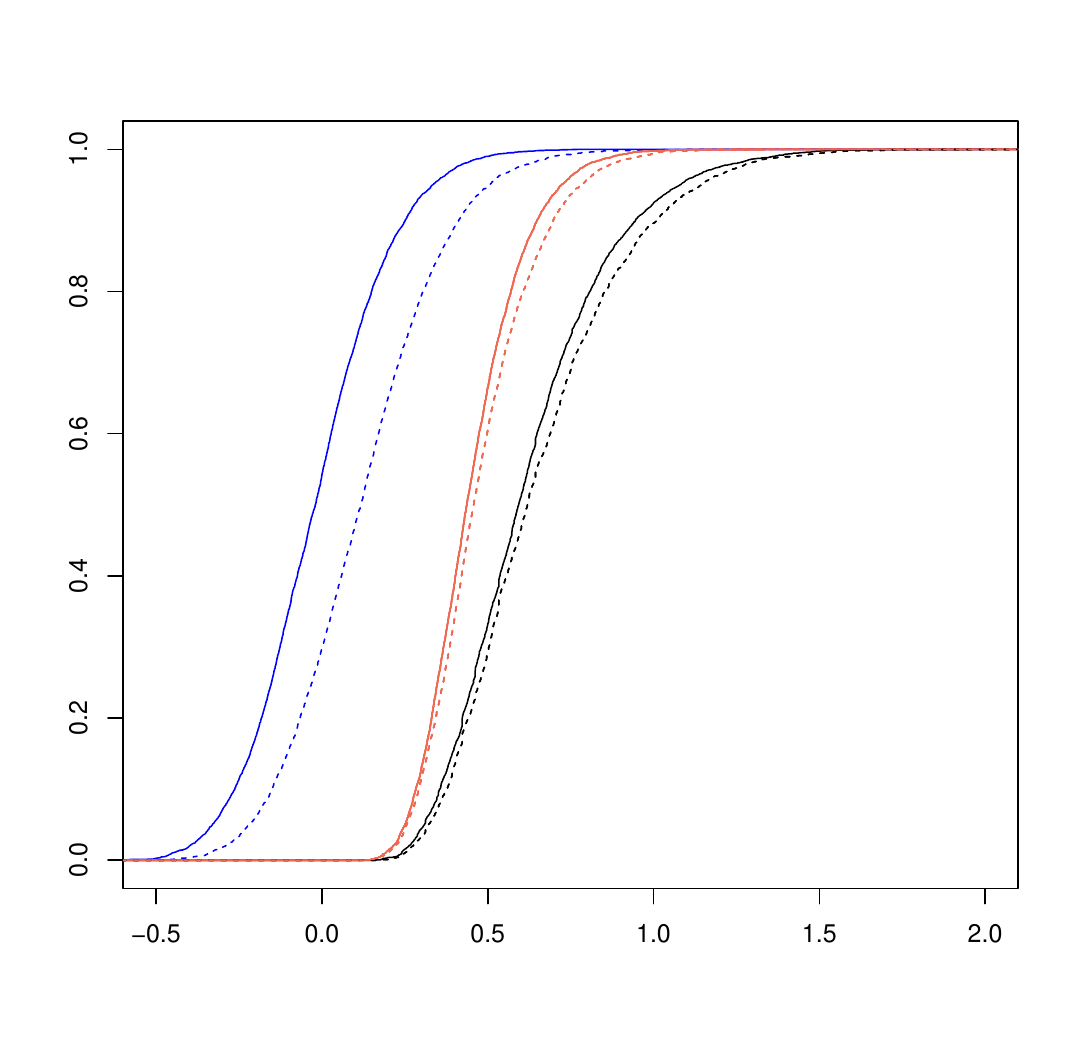}
\caption{Negative Binomial Alternative, with $k_{min}=3,k_{max}=12,p_{min}=0.082,p_{max}=0.053$; and $m_{nb}=70,p_{nb}=0.9091$.}
\label{erekl2}
\end{figure}

Similarly we investigate when the alternative is a negative binomial distribution.
The parameters are $m_{nb}=70$ and $p_{nb}=10/11$,
giving a mean value of 7 and a variance of 7.7.
The probabilities of the extreme cells mirror those in the first experiment, because we are still fitting a Poisson null with the same parameter.
Empirical distribution functions in Figure \ref{erekl2}
reinforce the results in Figure \ref{erekl1}: the $\hat c$ test performs more strongly than the gof tests in \eqref{e2}.

\section{Wider classes of alternative hypotheses}

\hspace{1.5em}
The beta-binomial distribution forms the basis of much statistical work for discrete data in the literature, and the next three scenarios tested are devoted to this case.  The results of these experiments appear in Figures \ref{erekl3}, \ref{erekl4} and \ref{erekl5}.

The fourth example, applying a particular mixture of binomial and negative binomial distributions, does not, to our knowledge, appear in the statistical literature, but is certainly feasible: it is not fanciful to imagine contaminated data, some of which is binomial and some of which negative binomial, with the records differentiating between the sample values lost.  Figure \ref{erekl6} corresponds to this experiment.

The final three examples of varying alternative hypotheses appear at first glance rather artificial, involving continuous distributions with observed values truncated to integers.  Denoting the continuous variable by $Z$, the observed discrete value $X$ has been rounded down to the lower integer: $X=\lfloor Z\rfloor$.  Results of these experiments with
the normal, gamma and Weibull distributions,
appear in Figures \ref{erekl7}, \ref{erekl8} and \ref{erekl9} respectively.

Again, while we are not aware of these final three scenarios having been investigated in the statistical literature, it is plausible that ostensibly count data arises in this way, especially when the origin of the data is unclear.
Rounding continuous data up or down to integral values is widespread, and not always acknowledged.

The first experiment of these three, for which $Z$ is normally distributed, is included because of the ubiquity of the normal distribution as an assumed genesis of data for statistical analysis.
In the present context, however, the second and third cases, involving the gamma and Weibull distibutions, have deeper posssible rationales as waiting distributions.

By analogy to some extent with the interpretation of the negative binomial as a waiting distribution for successes in a series of Bernoulli trials, a gamma distributed variable can be represented as the waiting time for events to arise in a Poisson process; and the Weibull distribution represents the waiting time for the next event when the intensity (the probability of occurrence in the next instant) may depend on the time elapsed since the last occurrence.
The gamma distribution has other representations, including the $\chi^2$ distribution; and the Weibull distribution is an essential element of statistical investigations in reliability theory.
Both distributions arise frequently in applied statistical analysis.

All seven experiments utilising alternative hypotheses from outside the Poisson/binomial/negative binomial family are consistent: the $\hat c$ test exhibits little power, and the MLE tests tend to outperform the gof tests using the assumed exact knowledge of the parameter.

\subsection{Beta-binomial Distribution}

\hspace{1.5em}
The beta-binomial distribution has found applications in several fields: see, {i.a.}, \cite{skellam1948a}, \cite{griffiths1973a},  \cite{wilcox1981a}, \cite{alanko-lemmens1996a}, \cite{chuang-stein1993a} and \cite{lao1997a}.
Its popularity is enhanced by  its convenient interpretation in a Bayesian context, since the beta and the binomial are conjugate distributions: see, for example, \cite{lee-sabavala1987a} and \cite{wilcox1977a}.

A draw from a beta distribution provides a value between 0 and 1, which then serves as the probability of success $p_{b}$ in an experiment of $m_{b}$ Bernoulli trials.  Shape parameters $\alpha\geq0$ and $\beta\geq0$ define the underlying beta distribution.

Setting $p_0=\frac\alpha{\alpha+\beta}$, the mean of the beta-binomial distribution is $m_{b}\,p_0$. Further letting $\rho=\frac1{\alpha+\beta+1}$, the variance is given by $m_{b}\,p_0(1-p_0)[1+(m_{b}-1)\rho]$.

The mean values in the three experiments are 5, 6 and 14, with parameter values shown in the captions to the figures.  In each case, the variance equals the mean.

\begin{figure}[htbp]\centering
  \includegraphics[width=\textwidth]{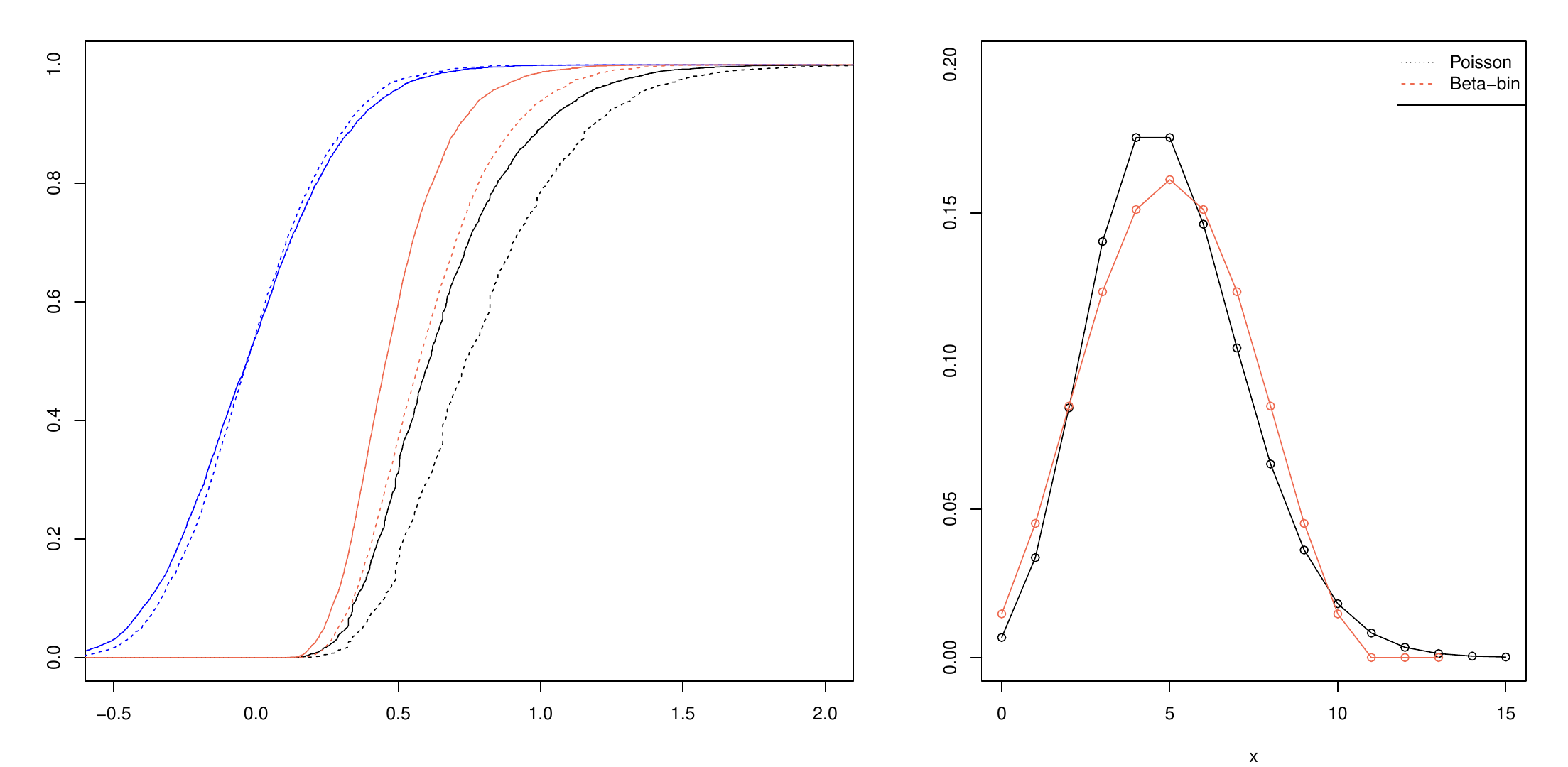}
  \caption{First Beta-binomial Alternative, with $k_{min}=1,k_{max}=9,p_{min}=0.040,p_{max}=0.068$; and $m_{b}=10,\alpha=4,\beta=4$.  Mean value is 5.}
\label{erekl3}
\end{figure}

\begin{figure}[htbp]\centering
    \includegraphics[width=\textwidth]{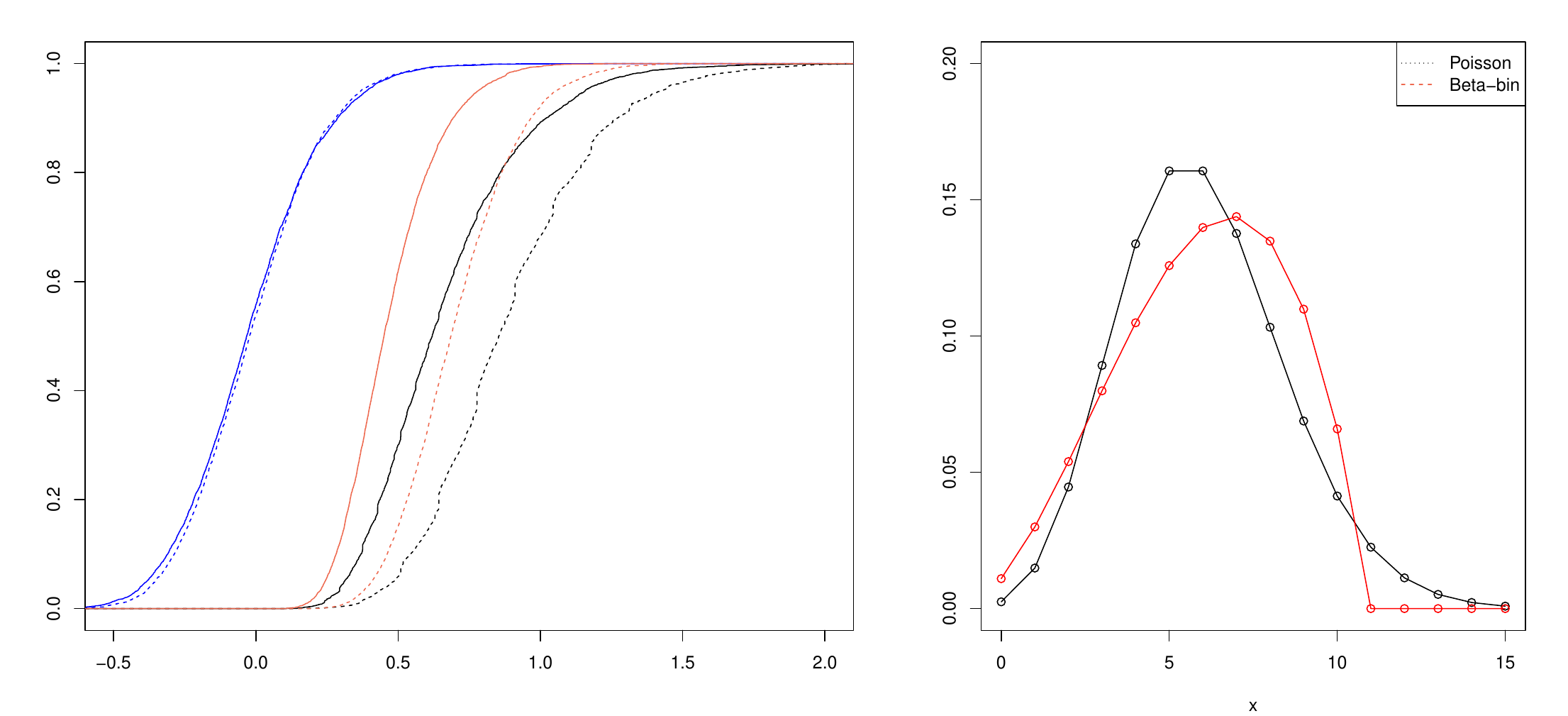}
  \caption{Second Beta-binomial Alternative, with $k_{min}=2,k_{max}=10,p_{min}=0.062,p_{max}=0.084$; and $m_{b}=10,\alpha=3,\beta=2$. Mean value is 6.}
\label{erekl4}
\end{figure}

\begin{figure}[htbp]\centering
  \includegraphics[width=\textwidth]{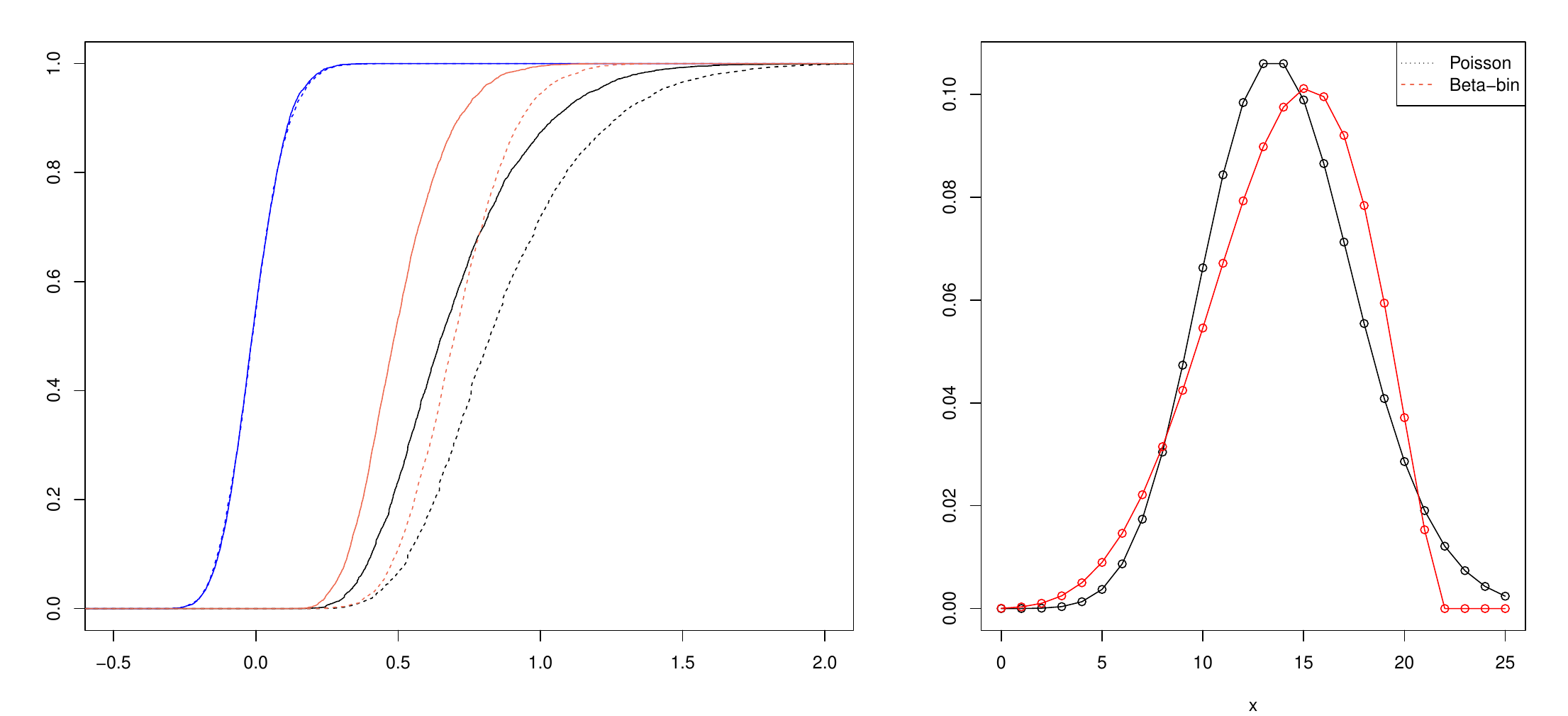}
  \caption{Third Beta-binomial Alternative, with $k_{min}=8,k_{max}=20,p_{min}=0.062,p_{max}=0.077$; and $m_{b}=21,\alpha=6,\beta=3$.  Mean value is 14.}
\label{erekl5}
\end{figure}

The EDFs comparing results from the Poisson first half of the experiment and the beta-binomial second half of the experiment are shown in the left hand side of Figures \ref{erekl3}, \ref{erekl4} and \ref{erekl5}; while the right sides of those Figures compare the corresponding pmfs.

All three plots comparing EDFs indicate that the $\hat c$ test has low power.
This is perhaps to be expected.  The $\hat c$ test is designed for use within the family of Poisson/binomial/negative binomial distributions, so that given alternative hypotheses with identical means and variances, the test provides no guidance as to whether the binomial or negative binomial might be preferred to a rejected Poisson distribution.

Figure \ref{erekl4} seems merely to reinforce the impression given by the other two figures plotting EDFs for the beta-binomial, but the gof test might not be so likely to be carried out.  A glance at the corresponding pmfs indicates a possible problem: the topmost cell in the gof test might well be empty.  For a large sample size, and fitting a Poisson with mean 6 (or close to it with the MLE), a suitable value for $k_{max}$ could be 11, with a corresponding $p_{max}=0.043$.  But the frequency in this cell is necessarily zero, because the maximum number of successes in the beta-binomial is $m_{b}=10$.  The final term in \eqref{e4} can be expected to be far larger than the other terms, and the gof test in \eqref{e2} to exhibit high power.  The problem is that an astute observer will notice the zero frequency in the topmost cell, and not bother testing this alternative distribution at all: it is plainly not a good fit to the data.  That said, an equally astute observer with a sample size of 50 will see no problem at all: setting $k_{max}=10$, there would be 4 observations expected and (on average) 3 observed in the topmost cell.

The same potential problem arises whenever the beta binomial distribution is considered as an alternative to the Poisson, since the latter has infinite support.  From Figures \ref{erekl3} and \ref{erekl5}, for instance, the same point can be made as in Figure \ref{erekl4}, but with less force:  in those cases, a shortage of observations in the final cell is not likely to be noticed before a formal gof test is carried out.

From \eqref{erekl5} we see that the gof tests using MLEs (continuous and broken lines in red) seem more powerful than the gof tests using the true parameter, shown in black: both tests use \eqref{e2}.

This is consistent with a claim by Khmaladze.
Referring to \eqref{e4}, he
explains that $Y(\hat\theta)$ is a projection of $Y(\theta)$ onto a contiguously shifted alternative hypothesis, and for distributional shifts orthogonal to the likelihood function in the underlying manifold,
the former has smaller stochastic variance than the latter.  The expected result is that
gof tests based on  $Y(\hat\theta)$ should have greater power than those based on $Y(\theta)$.  See \citet[\S2.3]{khmaladze2025a}.

The greater power of the MLE estimates is evident in Figures \ref{erekl6}, \ref{erekl8} and \ref{erekl9}, although less clearly present in Figure \ref{erekl7}.

\subsection{A mixture of binomial and negative binomial distributions}

\begin{center}\begin{figure}[htbp]
  \includegraphics[width=\textwidth]{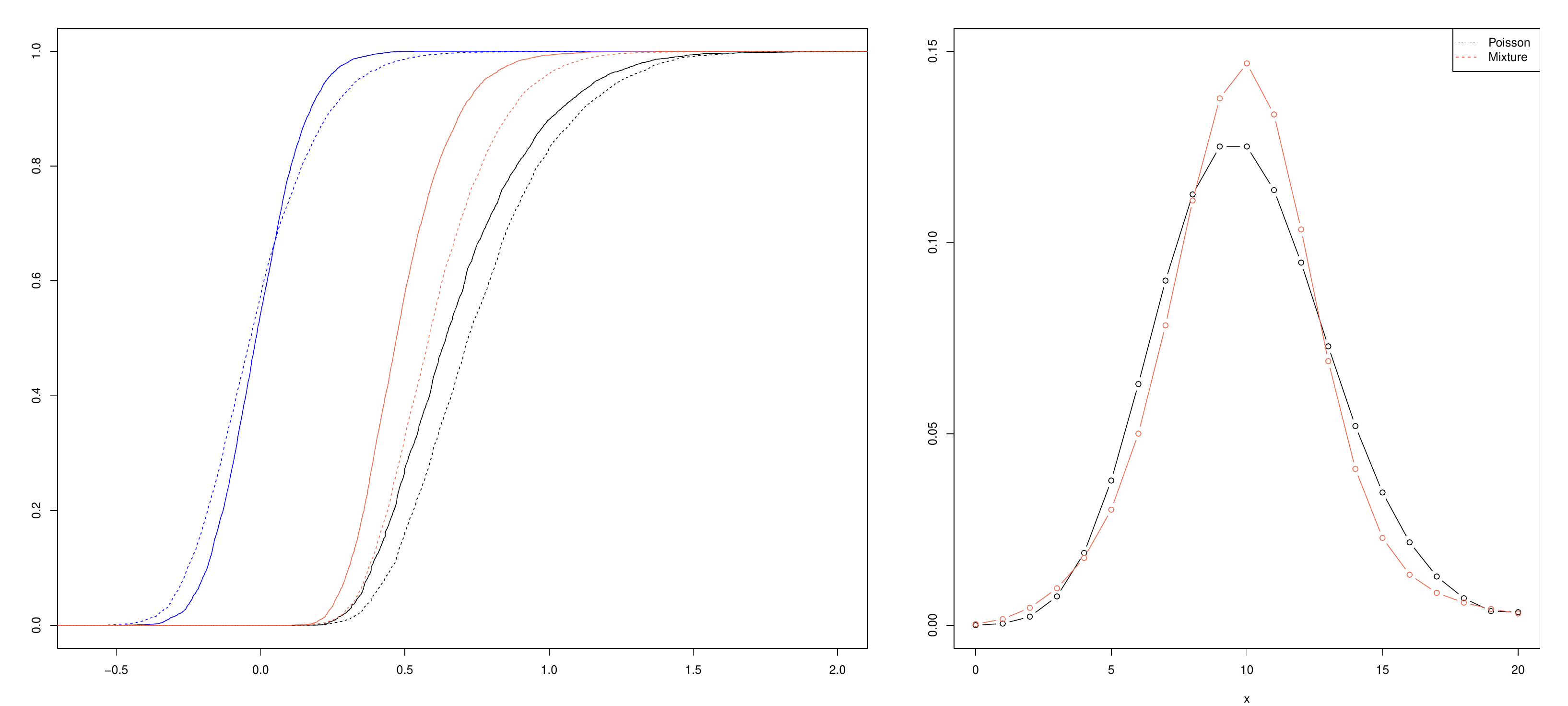}
  \caption{Mixture Alternative, with $k_{min}=5,k_{max}=15,p_{min}=0.067,p_{max}=0.083$; and $m_b=20,m_{nb}=10,p=0.5$.  Mean value is 10.}
  \label{erekl6}
\end{figure}\end{center}

Since the variance of a negative binomial random variate exceeds its mean, and the situation is reversed for the binomial, our next alternative hypothesis is a mixture of the two.  Each of the underlying distributions is chosen to have the same mean, and the mixing probability chosen so that the variance and mean of the mixture are identical.  As a result the $\hat c$ test has no reason to prefer the binomial or the negative binomial.

Symbolically, let $m_b=20$, $m_{nb}=10$, and the probability of success in each case be $p=1/2$.
The mean and variance of the binomial are $m_b\,p$ and $m_b\,p(1-p)$, while those for the negative binomial are $m_{nb}\,(1-p)/p$ and $m_{nb}\,(1-p)/p^2$.  With a weight of $w$ attached to the binomial distribution, the moments about zero are weighted averages of those of the constituent distributions.  Hence, denoting the variance of X by $VX$, and in an obvious notation, we have
\[
E\,X^2=VX+(E\,X)^2=w\ E\,X_b^2+(1-w)\,E\,X_{nb}^2
\]
Insertion of the values of the mean and variance of the constituent distributions leads quickly to the conclusion that $EX=VX=10$ when $w=2/3$.

The gof tests in Figure \ref{erekl6} clearly exhibit greater power than the $\hat c$ test, and the gof test applying $Y(\hat\theta)$ is seen to be more powerful than that utilising $Y(\theta)$, as noted in the discussion concerning Figure \ref{erekl5} above.

\subsection{A normal distribution}

\begin{figure}[htbp]\centering
  \includegraphics[width=\textwidth]{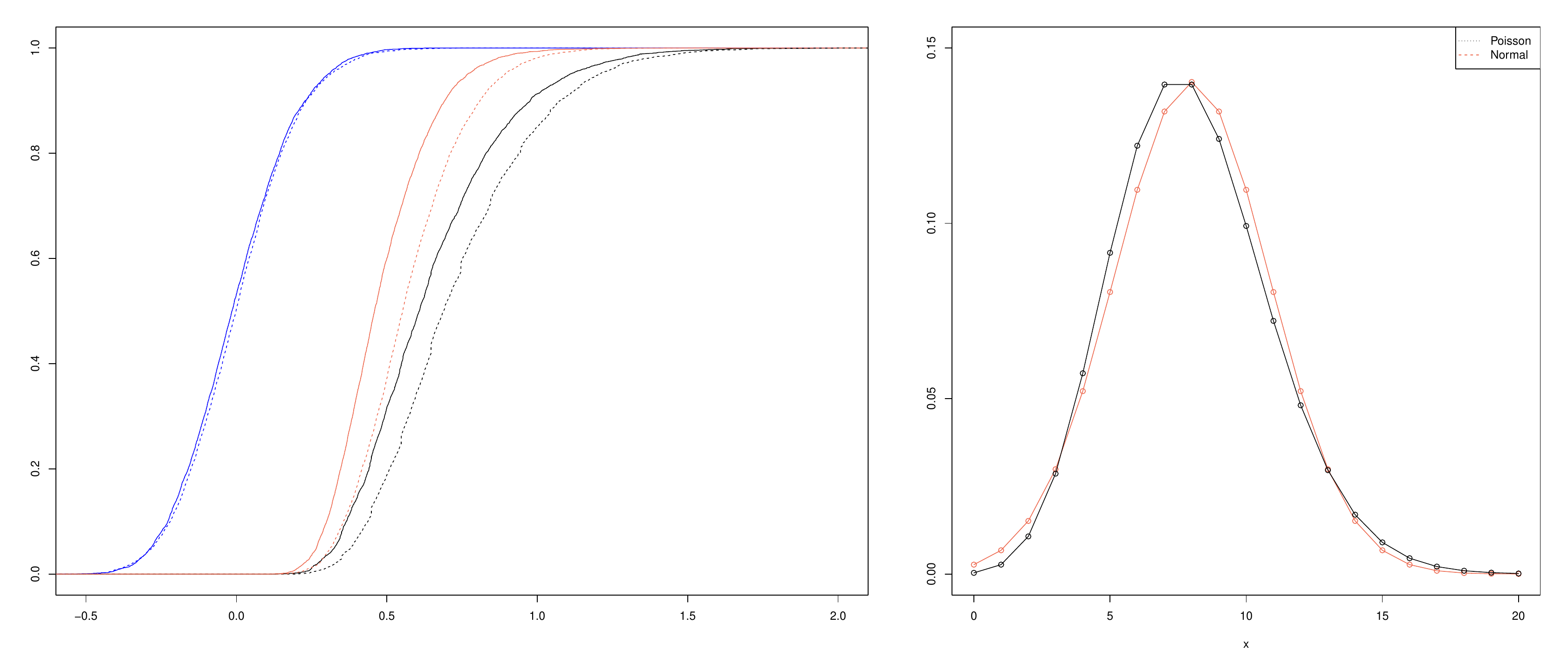}
    \caption{Normal Alternative, with $k_{min}=3,k_{max}=14,p_{min}=0.042,p_{max}=0.034$; and $a=8$.  Mean value is 8.}
\label{erekl7}
\end{figure}

In the first of the final three experiments, we 
let $Z\sim{\cal N}(a+0.5,a)$, and set $X=\lfloor Z\rfloor$, so that mean and variance for the discrete variable $X$ are almost equal for moderate $a>0$.  Repeating the experiment when the data $X$ is generated in this way is summarised in Figure \ref{erekl7}.  The principal message from other experiments is reinforced: the $\hat c$ test has little power against this alternative.
There appears some indication that the gof test utilising $Y(\hat\theta)$ is
more powerful than when $Y(\theta)$ is employed, but that message comes across less clearly than in the other experiments.

There is little point in allowing $a$ to become large in this experiment, because the Poisson distribution collapses to normality for large parameter values.

\subsection{Gamma and Weibull distributions}

\begin{figure}[htbp]\centering
  \includegraphics[width=\textwidth]{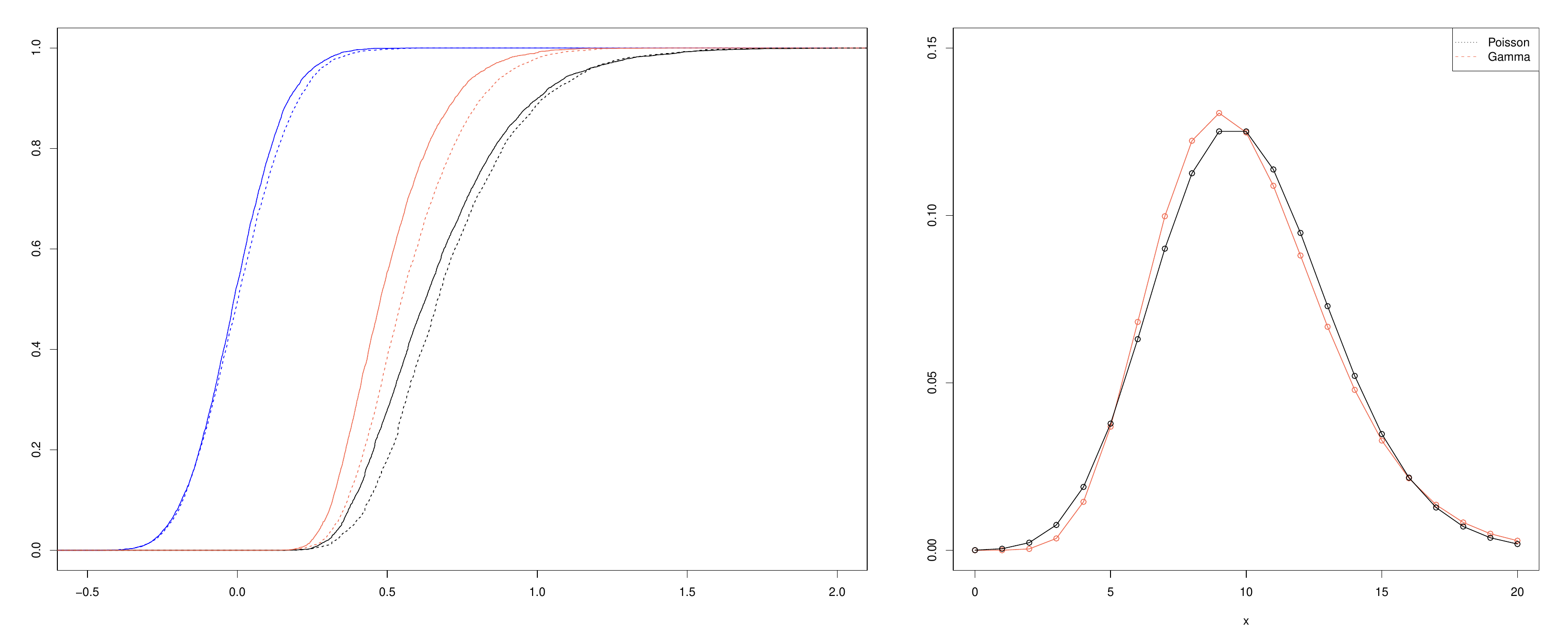}
    \caption{Gamma Alternative, with $k_{min}=4,k_{max}=17,p_{min}=0.029,p_{max}=0.027$; and $k=11.025,b=0.952$.  Mean value is 10.}
\label{erekl8}
\end{figure}

For the final two experiments we allow $Z$ to follow first the gamma distribution, then the Weibull distibution, again setting $X=\lfloor Z\rfloor$.

The probability density function (pdf) for the gamma distribution is proportional to $(z/b)^{k-1}\exp(-z/b)$, where $b>0$ is a scale parameter, and the so-called shape parameter $k$ is the number of events for which one is waiting (although in this experiment we do not insist that $k$ be an integer).

Disregarding a normalising constant, the pdf for the Weibull assumes the form $(z/b)^{k-1}\exp(-(z/b)^k)$. The parameter $b$ performs the same scaling function as for the gamma, while the parameter $k$ is still called a shape parameter, but with a different meaning.  See \cite{johnson-kotz-balakrishnan1994a}, {\it i.a.},
for detailed discussion of these distributions.

\begin{figure}[htbp]\centering
  \includegraphics[width=\textwidth]{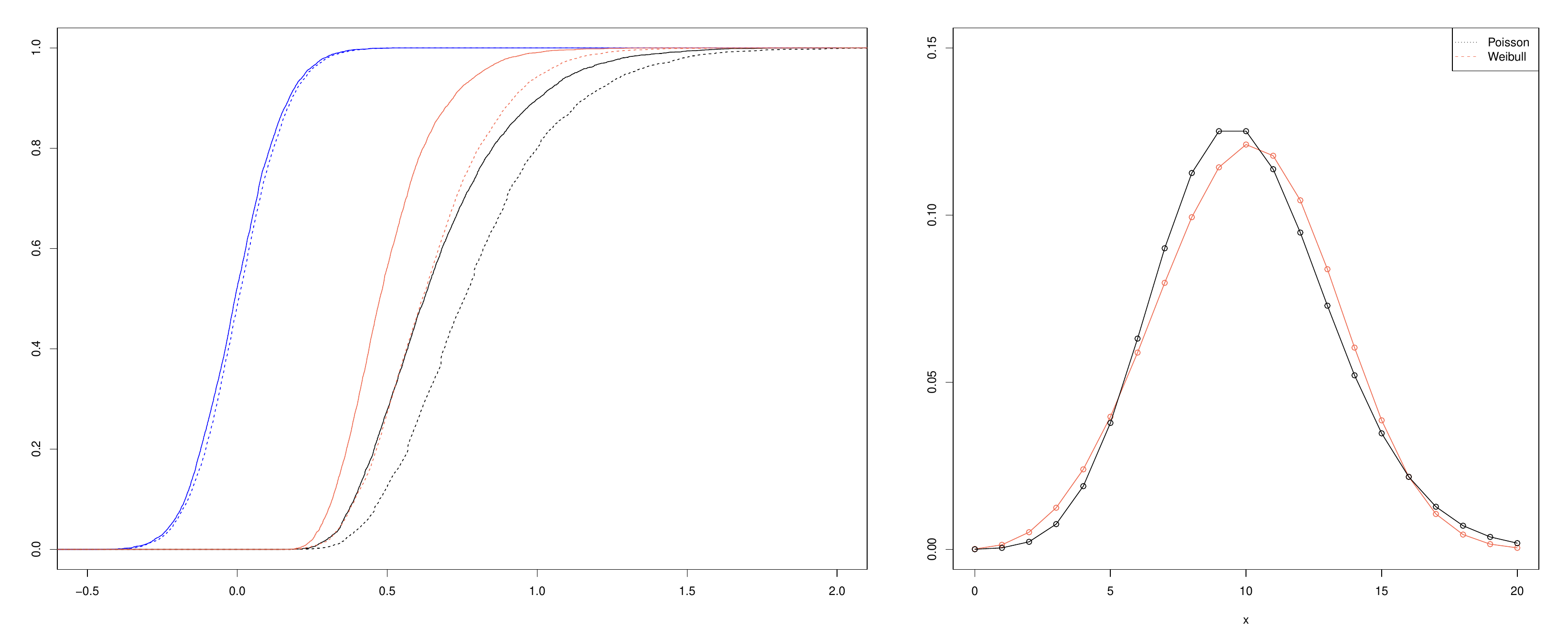}
    \caption{Weibull Alternative, with $k_{min}=4,k_{max}=17,p_{min}=0.029,p_{max}=0.027$; and $k=3.698,b=11.637$.  Mean value is 10.}
\label{erekl9}
\end{figure}

The values of $b$ and $k$ are listed in the captions to Figures \ref{erekl8} and \ref{erekl9}.  In both cases, the mean and variance of $X$ are equal to 10.

Once again we find the same messages reinforced: the $\hat c$ test has little power in comparison with the gof tests; and the gof test based on the MLE has greater power than that based on the true parameter.

\section{Conclusion}

\hspace{1.5em}Against alternative hypotheses of the binomial and negative binomial distributions, testing for a Poisson null hypothesis can have greater power than gof tests.  Faced with more general families of alternative distributions, however, at least for the $\hat c$ test used here, power to reject the null may be curtailed.

Our results also provide some evidence to support Khmaladze's claim that gof tests using maximum likelihood estimates may be more powerful than gof tests using, or guessing, the correct value of a parameter when testing for Poissonity.

We are not claiming the general superiority of the gof tests of the type considered in this paper over other possible tests.
Alternative hypotheses have been set up with the mean and variance equal to those of the null, making it difficult for any Poisson based test to disentangle the null and the alternative, especially since the Poisson distribution has but one parameter.
Under more general circumstances, against more `random' alternative hypotheses, tests based on $\hat c$ may perform well; but we have not investigated this possibility.

\end{document}